\documentclass[letterpaper, 10 pt, conference]{./cls/ieeeconf}  

\IEEEoverridecommandlockouts                              
\title{\LARGE \bf
Enhancements on a saturated control for stabilizing a quadcopter: adaptive and robustness analysis in the flat output space}

\author{Huu-Thinh Do$^{1}$ and Franco Blanchini$^2$ and Ionela Prodan$^{1}$
\thanks{$^{1}${Univ. Grenoble Alpes, Grenoble INP$^\dagger$, LCIS, 26000 Valence, France}.
Email: \textsf{ \{huu-thinh.do,ionela.prodan\}@lcis.grenoble-inp.fr}
\newline
$^\dagger$Institute of Engineering and Management Univ. Grenoble Alpes. 
}
\thanks{$^{2}$Dipartimento di Matematica e Informatica, Universit\`a di Udine, 33100 Udine, Italy. Email: \textsf{franco.blanchini@uniud.it}}
\thanks{This work is partially funded by La Région, Pack Ambition Recherche 2021 - PlanMAV and by the French National Research Agency in the framework of the ``Investissements d'avenir" program ``ANR-15-IDEX-02" and the LabEx PERSYVAL ``ANR-11-LABX-0025-01".
Franco Blanchini has been supported by European Union, NextGeneration EU
(Grant Uniud-DM737)}
}
\usepackage[hidelinks]{hyperref} 
\hypersetup{
    pdftitle={FASC HTD\_FB\_IP}
    }
\newcommand{\be}{\begin{equation}}
\newcommand{\ee}{\end{equation}}

\newcommand{\bald}{\begin{aligned}}
\newcommand{\eald}{\end{aligned}}

\newcommand{\bbm}{\begin{bmatrix}}
\newcommand{\ebm}{\end{bmatrix}}

\newcommand{\bxi}{\boldsymbol{\xi}}
\newcommand{\bu}{\boldsymbol{u}}

\newcommand{\bx}{\boldsymbol{x}}
\newcommand{\bv}{\boldsymbol{v}}
\newcommand{\bI}{\boldsymbol{I}}
\newcommand{\R}{\mathbb{R}}

\newcommand{\umax}[1]{{#1}_{max}}

\newtheorem{rem}{Remark}
\newtheorem{prop}{Proposition}

\newcommand{\QEDA}{\hfill\ensuremath{\square}}
\newcommand{\qed}{\hfill\blacksquare}

\usepackage{tikz}
\usepackage{physics}
\usepackage{amsmath}
\usepackage{tikz}
\usepackage{mathdots}
\usepackage{yhmath}
\usepackage{cancel}
\usepackage{color}
\usepackage{siunitx}
\usepackage{array}
\usepackage{multirow}
\usepackage{amssymb}
\usepackage{tabularx}
\usepackage{extarrows}
\usepackage{booktabs}
\usetikzlibrary{fadings}
\usetikzlibrary{patterns}
\usetikzlibrary{shadows.blur}
\usetikzlibrary{shapes}

\usepackage{pgfplots}
\pgfplotsset{width=10cm,compat=1.9}

\usepackage{cases} 
\usepackage[ruled,norelsize,linesnumbered]{algorithm2e}
\usepackage{cite} 
 \usepackage{mathtools} 

\newcommand{\bw}{\boldsymbol{w}}
\newcommand{\satlambda}{\text{sat}_\lambda (-\gamma \boldsymbol B^\top \boldsymbol P\bxi)}
\newcommand{\row}{\text{r}}

\newtheorem{definition}{Definition}
\begin{document}

\maketitle
\thispagestyle{empty}
\pagestyle{empty}

\begin{abstract}
This paper extends our previous study on an explicit saturated control for a quadcopter, which ensures both constraint satisfaction and stability thanks to the linear representation of the system in the flat output space. The novelty here resides in the adaptivity of the controller's gain to enhance the system's performance without exciting its parasitic dynamics and avoid lavishing the input actuation with excessively high gain parameters. 
Moreover, we provide a thorough robustness analysis of the proposed controller when additive disturbances are affecting the system behavior.
Finally,
simulation and experimental tests validate the proposed controller.
\end{abstract}
\section{Introduction}
In the literature, quadcopters have always been of special interest
due to their wide spectrum of applicability. Thus, the vehicle's navigation and control are commonly addressed. 
Typically, on one hand, the trigonometrical complexity of the system is governed by employing online optimization-based control so that both physical constraints and stability are ensured. For example, a non-linear model predictive control (MPC) technique was presented in \cite{nguyen2020stability,nguyen2019stabilizing}, with the two requirements guaranteed by the existence of a local controller together with the necessary terminal ingredients. In \cite{lee2010geometric}, a coordinate-free control approach was presented which ensures the attractiveness while avoiding both singularities and ambiguities of the Euler and quaternion representations. However, the above-mentioned solutions, while reliable, come with a significant computational cost due to the online-solving of the implicit control law, a sophisticated synthesis or an insufficient concern on the system's physical limitation.

On the other hand, falling in the class of \textit{differentially flat systems}\cite{levine2009analysis}, the quadcopter's dynamics are compliant to the equivalent linear integrator chains, in the new coordinates of \textit{the flat output space}.
The advantage of the linear dynamics was investigated in various control designs, \cite{nguyen2020flat,mellinger2011minimum}.
However, this approach, as well as other techniques employing feedback linearization, encounters the problem of convoluted constraints in the new coordinates and usually deals with conservative subsets of the feasible domain \cite{limaverde2016trajectory}.

To account for the previously mentioned conservativeness and to further advocate the computational advantage of the system's linear representation in the flat output space, in \cite{do2023flatness}, we proposed an explicit saturation function particularized for a quadcopter's translational dynamics to ensure constraint satisfaction. 
It is also noteworthy that this convoluted constraint set in the flat output space stands out from the classical amplitude bound constraints discussed in the literature \cite{hu2002analysis,hu2001control,nguyen2018flat}. 
Then, in combination with a nominal control design based on a pre-stabilization procedure, a saturated control was established by scaling the nominal gain by a scalar factor, $\gamma$,
with the only condition of being $\gamma\geq 1$.
Theoretically, a gain scaled by $\gamma\geq 1$ can have an arbitrarily high value without changing the stability and constraint satisfaction guarantees. However, practically, excessively high-gain feedback may cause either the excitement of the non-modeled neglected dynamics (e.g., rotational dynamics, propeller actuation) or control over-actuation.
Therefore, to better select $\gamma$, we enhance the proposed controller through a so-called $\lambda$-tracking adaptation \cite{blanchini2002adaptive,ilchmann1994universal}. Lastly, to complete the scheme, we adjust the synthesis 
when external disturbances are taken into account and when the control objective is changed from stabilization to trajectory tracking. Finally, the adaptive scheme is validated via experiments. 
Briefly, our contributions are summarized in the following. We:
\begin{itemize}
    \item establish an adaptive scheme to select a suitable gain for the explicit saturated control and provide upper bounds for the non-decreasing convergence of this factor;
    \item analyze the robustness of the control synthesis when the system is affected by bounded external disturbances;
    \item validate and examine the theoretical results via simulation and experimental tests for a nano-drone hovering and multiple drones reference tracking. The experiment video can be found at: 
    \href{https://youtu.be/kwfXiJ6odnc}{https://youtu.be/kwfXiJ6odnc}.
\end{itemize}

The remainder of the paper is structured as follows. Section \ref{sec:sys_pre} recalls the system description and the linearizing transformation for the quadcopter. Also, some prerequisites for the saturated control design are delineated. Section \ref{sec:adapt} constructs the adaptive scheme for a proper gain's selection and characterizes the robustness of the proposed controller.
Experimental validation is provided in Section \ref{sec:expval}. Finally, Section \ref{sec:conclude} draws the conclusion and discusses future works.

\textit{Notation:} Bold capital letters refers to matrices with appropriate dimension. For a matrix $\boldsymbol B$, $\row_i(\boldsymbol B)$ denotes its $i$-th row. $\boldsymbol Q\succ 0,\boldsymbol Q\succeq 0$ implies $\boldsymbol Q$ is positive definite and positive semi-definite, respectively. $\boldsymbol P\prec 0\, (\boldsymbol P \preceq 0)$ implies $-\boldsymbol P\succ 0 \,(-\boldsymbol P \succeq 0)$.
The time variable $t$ will be considered when necessary. 
Bold lower-case letters denote vectors and for $\boldsymbol P\succ 0$, $\|\bx\|_{\boldsymbol{P}}\triangleq\sqrt{\bx^\top \boldsymbol P \bx}$, $\|\bx\|_2\triangleq \sqrt{\bx^\top \bx}$.
$\boldsymbol 0_{m\times n}$ is a zero matrix of size ${m\times n}$ ,
$\boldsymbol 0_n\triangleq\boldsymbol 0_{n\times n}$. $\bI_n$ is the identity matrix of size $n$.
$\ominus$ denotes the Pontryagin difference.
\section{System description and preliminaries}
\label{sec:sys_pre}
\subsection{Quadcopter's LTI representation in closed-loop}
Let us recap the system reformulation via feedback linearization for a quadcopter.
Firstly, the translational dynamics of the thrust-propelled vehicle are expressed as:
\begin{align}
& \dot \bxi = \boldsymbol A\bxi + \boldsymbol B(\boldsymbol R_\psi \boldsymbol f(\bu) - g\boldsymbol{e}_3),\nonumber\\
 &\boldsymbol A=\bbm \boldsymbol{0}_{3}\; \bI_{3} \\
\boldsymbol{0}_{3}\;\boldsymbol{0}_{3}\ebm ,
\boldsymbol B=\bbm \boldsymbol{0}_{3}\\ \bI_{3}\ebm,\bxi=\bbm \boldsymbol p \\ \dot{\boldsymbol  p} \ebm, \bu=[T , \phi, \theta]^\top , \label{eq:nonlinear_drone} \\
&\boldsymbol R_\psi
\triangleq \bbm\cos\psi & \sin \psi & 0 
\\ \sin\psi & -\cos\psi & 0 \\ 
0 & 0 & 1\ebm,
\boldsymbol f(\bu)\triangleq T
\bbm 
		\cos \phi \sin \theta \\
		\sin \phi\\
		\cos \phi \cos \theta 
	\ebm 
 \nonumber,
\end{align}
where $\boldsymbol p\triangleq[x,y,z]^\top$ (m) collects the drone's position in the global frame, $\boldsymbol e_3\triangleq[0,0,1]^\top$ represents the basis vector along the $z$ axis, vector $\bu\in\R^3$ denotes the input including the normalized thrust $T$, the roll $\phi$ and the pitch angle $\theta$, respectively. $\psi $ denotes the measured yaw angle and $g $ is the gravity acceleration. Furthermore, the actuation constraints of the input $\bu$ are described by a set $\mathcal{U}$ as:
\be 
\bu\in\mathcal{U}\triangleq\{\bu|0\leq T\leq \umax T;|\phi| \leq \umax\epsilon; |\theta| \leq \umax\epsilon \}.
\label{eq:constrU}
\ee 
where $\umax T>0$ and $0<\umax \epsilon<\pi/2$ are, respectively, the constant maximum thrust and inclination of the vehicle.

Then, by denoting $\bv=[v_1,v_2,v_3]^\top\triangleq \boldsymbol R_\psi \boldsymbol f(\bu) - g\boldsymbol e_3$, we can deduce the linearizing variable change as:
\be 
\bu=\boldsymbol{\Phi}(\bv)\triangleq\boldsymbol f^{-1}(\boldsymbol R_\psi^{-1}(\bv+g\boldsymbol e_3)),
\label{eq:reformU}
\ee 
where, for $\boldsymbol h=[h_1,h_2,h_3]^\top\in\R^3$, the function $\boldsymbol f^{-1}(\boldsymbol h)$ is defined as \cite{thinhECC23}:
\be 
\boldsymbol f^{-1}(\boldsymbol h)\triangleq [\|\boldsymbol h\|_2,\, \sin^{\text{-1}}(h_2/\|\boldsymbol h\|_2),\,\tan^{\text{-1}}(h_1/h_3)]^\top.
\ee

Consequently, under the condition $v_3\geq -g$,
the model \eqref{eq:nonlinear_drone} can be transformed into the linear system:
\be 
\dot \bxi = \boldsymbol A\bxi + \boldsymbol B\bv,
\label{eq:drone_linear}
\ee 
with the reformulation of the input $\bu=\boldsymbol{\Phi}(\bv)$ as in \eqref{eq:reformU}.
Similarly, the input constraint set $\mathcal{U}$ is reformed into:
\be 
\bv\in\mathcal{V}=\{\bv\in\R^3:\boldsymbol{\Phi}(\bv)\in\mathcal{U}\}.
\ee 
It was shown in \cite{thinhECC23}, that $\boldsymbol{\Phi}(\bv)$ can be deduced directly from the {flat representation} of the system. Namely, all the system's variables can be expressed algebraically in terms of a special output, called \textit{flat output}, which is chosen as $\boldsymbol p$ in this system. Moreover, in the new coordinates (called \textit{the flat output space}), the set $\mathcal{V}$ is non-convex and time-variant ($\psi$-dependent), which proves challenging for both control design and real-time implementation. Thus, a compact convex subset $\mathcal{V}_c$ of $\mathcal{V}$ is alternatively employed as \cite{thinhECC23,freire2023flatness}:
\begin{align}
&\bv\in\mathcal{V}_c\triangleq\Big\{\bv=[v_1,v_2,v_3]^\top:\|\bv+g\boldsymbol e_3\|_2^2\leq \umax T^2; \nonumber\\ 
& (v_1^2+v_2^2)/\tan^2\umax \epsilon\leq(v_3+g)^2;v_3\geq -g \Big\}. \label{eq:Vc_set}
\end{align}
Hereinafter, we will employ $\mathcal{V}_c$ as the constraint set for the new variable $\bv$, which implies $\bu\in\mathcal{U}$ as in \eqref{eq:constrU}.
\subsection{Explicit saturated control for the quadcopter}
In this section, we summarize our previous work and recall the invariance conditions under disturbances. 

\begin{definition} \textit{(Explicit saturation function for $\mathcal{V}_c$ \cite{do2023flatness})}

Let us define a saturation function ensuring the constraint $\bv\in\mathcal{V}_c,\forall \bv\in\R^3$ given in \eqref{eq:Vc_set} as the following:
    \be
         \text{sat}_\lambda (\bv) \text{=}\lambda^*(\bv)\bv ,        
\text{with } \lambda^*(\bv)\text{=}
 \begin{cases}
            1 &\text{ if } \bv\in \mathcal{V}_c,\\
 \underset{{
\lambda \bv \in \mathcal{V}_c}}{ \max }\lambda
 &\text{ if } \bv\notin \mathcal{V}_c.
         \end{cases}
    \label{eq:def_sat_full}
\ee
Furthermore, when $\bv\notin \mathcal{V}_c$, the saturation factor $ \lambda^*(\bv)$ can be explicitly found by \cite{do2023flatness}:
\be
\lambda^*(\bv) = \max \mathcal{L}(\bv)\cap (0;1],
\label{eq:explicit_L}
\ee 
where, for $i\in\{1,2\}$,
\be
 \mathcal{L}(\bv)\triangleq\left\{ \dfrac{-g}{v_3}, \dfrac{a_0-g}{v_3},\dfrac{-b_i\pm \sqrt{b_i^2-4a_ic_i}}{2a_i}\right\}
\label{eq:KKT_M}
\ee 
$$
\begin{cases}
a_0=T_{max}\cos\umax\epsilon,a_1 = v_1^2+v_2^2-v_3^2\tan^2\epsilon_{max},&\\
b_1 =-2\tan^2\epsilon_{max}v_3g,c_1={-}\tan^2\epsilon_{max}g^2, &\\
a_2=v_1^2+v_2^2+v_3^2,\,b_2=2v_3g,c_2=g^2-T_{max}^2.&
\end{cases}
$$
\end{definition}
The geometric interpretation of the vector $\lambda^*(\bv) \bv $
is the
longest vector in $\mathcal{V}_c$ sharing the same direction with $\bv$. Next, recall the synthesis for a saturation control as follows.

\begin{prop}
\label{prop:saturatedcontrol}
Suppose there exists $\rho>0$ such that:
\be 
\mathcal{P}=\{\bv\in\R^3:\|\bv\|_2^2\leq \rho\}\subset \mathcal{V}_c \text{ as in \eqref{eq:Vc_set}}.
\label{eq:BallinVc}
\ee 
Then, 
$\forall \gamma\geq 1$,
under the gradient-based control \cite{blanchini2008set}:
\be 
\bv = \text{sat}_\lambda(-\gamma \boldsymbol B^\top \boldsymbol P \bxi),
\label{eq:controller_sat}
\ee 
the following ellipsoidal set is rendered invariant:
\be 
\mathcal{E}  =\{\bxi\in\R^6:\bxi^\top \boldsymbol P \bxi\leq \varepsilon\},
\label{eq:ellipsoidE}
\ee 
with
\begin{subequations}
\label{eq:find_ellipsoid}
\begin{align}
&\varepsilon^*={ \max }\;\varepsilon 
\label{eq:find_ellipsoid_a}\\
\text{s.t. }
\bbm
\boldsymbol P  & \boldsymbol 0_{6 \times 1} \\ \boldsymbol 0_{1 \times 6} & -\varepsilon
\ebm 
-&\tau
\bbm
\boldsymbol P \boldsymbol  B  \boldsymbol B^\top\boldsymbol  P  & \boldsymbol 0_{6 \times 1} \\ \boldsymbol  0_{1 \times 6} & -\rho 
\ebm \succeq 0
, \tau \geq 0,
\label{eq:find_ellipsoid_b}
\end{align}
\end{subequations}
if for some scalar $\alpha>0$, $\boldsymbol Q\triangleq \boldsymbol P^{-1}\succ 0$ satisfies the LMI:
\be 
\boldsymbol Q\boldsymbol A^\top +\boldsymbol A\boldsymbol Q-2\boldsymbol B\boldsymbol B^\top \preceq - \alpha \boldsymbol Q.
\label{eq:stabilizingcondi}
\ee 
Moreover, the closed-loop system:
\begin{equation}
    \dot\bxi=\boldsymbol A\bxi+\boldsymbol B\satlambda
\end{equation}
attains exponential stability
inside
$\mathcal{E} $.\QEDA
\end{prop}

\textit{Sketch of proof:}
The invariance of $\mathcal{E}$ as in \eqref{eq:ellipsoidE} can be briefly proven by showing that the Lyapunov function:
\be
V(\bxi)\triangleq\bxi^\top \boldsymbol P \bxi 
\label{eq:LyaFunc}
\ee 
has a non-positive time-derivative if $\bxi\in\mathcal{E}$. 
Firstly, by solving the  linear matrix inequality (LMI) \eqref{eq:stabilizingcondi} and then the problem \eqref{eq:find_ellipsoid}, the following implication will hold:
\be 
\text{if } \bxi^\top \boldsymbol P \bxi \leq \varepsilon \Rightarrow \|\boldsymbol B^\top \boldsymbol P\bxi\|^2\leq\rho,
\label{eq:condition_inputball}
\ee 
since condition \eqref{eq:find_ellipsoid_b} describes the equivalence of  \eqref{eq:condition_inputball} via an $\mathcal{S}$-procedure. Thus, together with the condition $\mathcal{P}\subset \mathcal{V}_c $ and the definition \eqref{eq:def_sat_full}, we can state:
\be 
\begin{aligned}
\|-\boldsymbol B^\top \boldsymbol P\bxi\|^2_2\leq \rho \leq \|\text{sat}_\lambda(-\gamma \boldsymbol B^\top \boldsymbol P \bxi)\|_2^2, \forall \bxi \in \mathcal{E},&\\
\Leftrightarrow\gamma\lambda^*(-\gamma \boldsymbol B^\top \boldsymbol P \bxi) \geq 1,\forall \bxi \in \mathcal{E}.&
\end{aligned}
\label{eq:conditionlambdagamma}
\ee 
Consequently, from \eqref{eq:conditionlambdagamma} and \eqref{eq:stabilizingcondi}, we can show that:
\be
\dot V(\bxi)=2\bxi ^\top\boldsymbol P (\boldsymbol A\bxi+\boldsymbol B\satlambda)\leq -\alpha V(\bxi).
\label{eq:converge_V}
\ee 


    
Thus, the proof is completed, while detailed computation and the maximum $\rho$ in \eqref{eq:BallinVc} can be found in \cite{do2023flatness}. $\qed$
\begin{prop}
\label{prop:invariance_robust}
    (\textit{Invariance condition for autonomous disturbed system \cite{blanchini2008set}:})
Consider the autonomous system affected by additive disturbances:
\be 
\dot \bxi =\boldsymbol A_{cl}\bxi + \boldsymbol E\bw,
\label{eq:invariance_cond_sys}
\ee 
with $\bw$ denoting the disturbance constrained as $\bw^\top \bw \leq 1$. Then, the ellipsoid $\mathcal{E}_w\triangleq\{\bx:\bx^\top \boldsymbol Q_w^{-1}\bx\leq 1\}$ is invariant under the dynamics \eqref{eq:invariance_cond_sys} if, for some $\beta>0$:
\be 
\boldsymbol Q_w\boldsymbol A_{cl}^\top +\boldsymbol A_{cl}\boldsymbol Q_w + \beta  \boldsymbol Q_w + \beta^{-1}\boldsymbol E\boldsymbol E^\top\preceq 0 .
\ee 
\end{prop}

\section{Adaptive saturated control \texorpdfstring{\\for the quadcopter and robustness analysis}{}}
\label{sec:adapt}

As previously mentioned, although the stability can be maintained for all $\gamma\geq 1$, the gain should not be chosen excessively high. In the following, let us discuss an adaptive law assigned to $\gamma$ to remedy the problem of gain selection.

\subsection{Saturated control with adaptive gain}
As noted above, exploiting the controller's stabilizing effect for all $\gamma\geq 1$, we assign to $\gamma(t)$ a non-decreasing dynamics and a stopping condition associated with the convergence of the Lyapunov function \eqref{eq:LyaFunc}. In such manner, $\gamma(t)$ can be proven to converge to a finite bound, solving the problem of manually tuning the parameter. More specifically, consider the following control mechanism:
\begin{subequations}
 {\label{eq:adapt_gamma}}
    \begin{align}
    \bv &= \text{sat}_\lambda(-\gamma(t) \boldsymbol B^\top \boldsymbol P \bxi), \label{eq:adapt_gamma_a}\\ 
    \dot \gamma  &=\mu\sigma(V(\bxi)),  \,
    \gamma(0)= \gamma_0\geq 1,\label{eq:adapt_gamma_b}
\end{align}
\end{subequations}
where, $\sigma(\mathrm{s})$ defines the threshold function\cite{ilchmann1994universal,blanchini2002adaptive}:
\be 
\sigma(\mathrm{s})=
\begin{cases}
    0 & \text{ if } 0\leq \mathrm{s}\leq V_\infty , \\
    \mathrm{s} - V_\infty &\text{ if } \mathrm{s}\geq V_\infty,
\end{cases}
\label{eq:thresholdV}
\ee 
while $\mu>0$ is the scalar factor adjusting the adaptation speed and $V_\infty$ is the stopping threshold for the adaptation of $\gamma(t)$. The matrix $\boldsymbol P$ is designed as in Proposition \ref{prop:saturatedcontrol} with the invariant set $\mathcal{E}$ as in \eqref{eq:ellipsoidE}.
\begin{rem}
As discussed in \cite{blanchini2002adaptive}, with the adaptive law \eqref{eq:adapt_gamma}, $\gamma(t)$ describes a non-decreasing value function starting from $\gamma(0)=\gamma_0\geq 1$, Thus, by the synthesis in Proposition \ref{prop:saturatedcontrol}, the stability and invariance of the system remain valid inside the ellipsoid $\mathcal{E}$ as in \eqref{eq:ellipsoidE}. 
Meanwhile, it can be proven that $\gamma(t)$ when $t\rightarrow\infty$ is bounded by a finite scalar. More specifically, from a given initial point $\bxi(0)\in \mathcal{E}$, when $V(\bxi)\geq V_\infty$, owing to \eqref{eq:converge_V}, we have:
\be 
\begin{aligned}
    {d}\gamma(t)/{dt}&=\mu (V(\bxi(t)) -V_\infty  )\leq \mu (V(\bxi(0))e^{-\alpha t} -V_\infty  )\\
\Rightarrow  \gamma(t_\infty)  &\leq  \gamma(0) 
+ \mu(
V_\infty  t_\infty + {\alpha^{-1}}V(\bxi(0))(1-e^{-\alpha t_\infty})
) 
\end{aligned}
\label{eq:bound_gamma}
\ee 
where $t_\infty$ denotes the required time for $V(\bxi(t))$ to reach $V_\infty$ starting from $V(\bxi(0))$ which can be bounded from the exponential convergence of $V(\bxi)$ (given in \eqref{eq:converge_V}) as:
\be 
\begin{aligned}
   {d}V(\bxi)/{dt}&\leq -\alpha V(\bxi)  \\
   \Leftrightarrow \ln{\dfrac{V_\infty}{V(\bxi(0))}}&\leq -\alpha t_\infty 
   \Leftrightarrow t_\infty
   \leq \dfrac{1}{\alpha}\ln{\dfrac{V(\bxi(0))}{ V_\infty}}.
\end{aligned}
\label{eq:bound_tinfty}
\ee 
Hence, from \eqref{eq:bound_gamma} and \eqref{eq:bound_tinfty}, we can compute the finite upper bound of $\gamma(t)$.
\end{rem}


\emph{Simulation study:}
For the sake of illustration and parameter study, let us simulate the proposed controller in the following setup. We study the behavior of the system \eqref{eq:nonlinear_drone} under the controller \eqref{eq:adapt_gamma} from the initial state: 
   $\bxi(0)=[       
    1.05,
    1.04,
    0.85,
   -0.02,
   -1.39,
   -0.42,
]^\top $ 
   where the adaptive gain $\mu$ is examined with different values. 
Detailed numerical parameters are provided in TABLE \ref{tab:para_sim_adpt}.
\begin{figure}[htbp]
    \centering
    \includegraphics[width=0.45\textwidth]{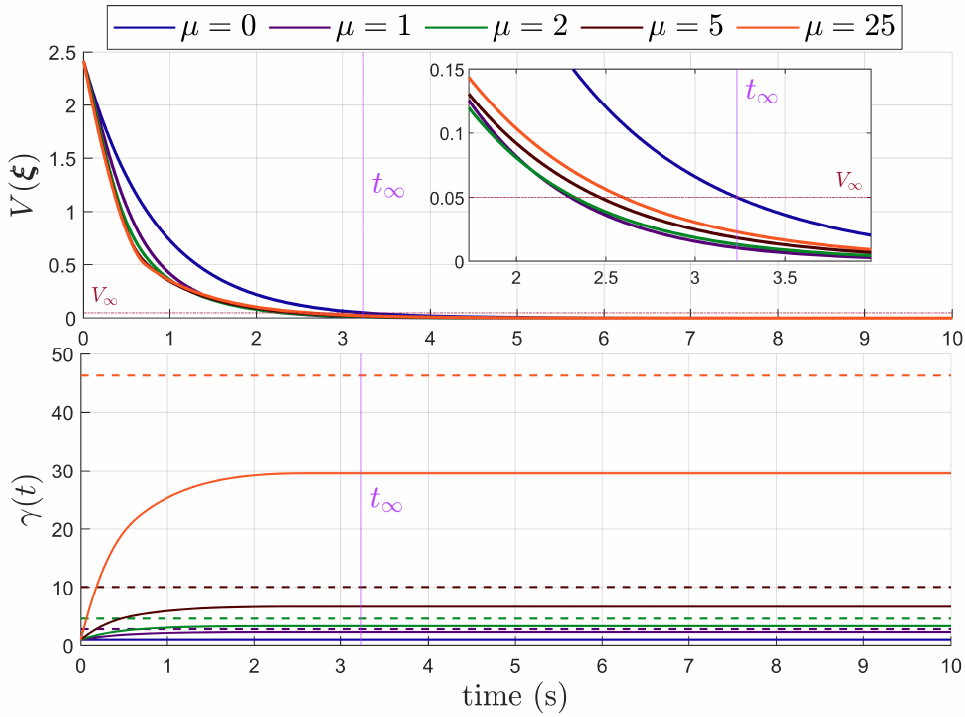}
    \caption{Lyapunov function $V(\bxi)$ (top), the evolution of $\gamma(t)$ and its upper bound, which is color-coded with $\mu$ in dashed-line (bottom).}
    \label{fig:adaptive_sim}
    \vspace{-0.5cm}
\end{figure}

\begin{figure}[htb]
    \centering
    \includegraphics[scale=0.46]{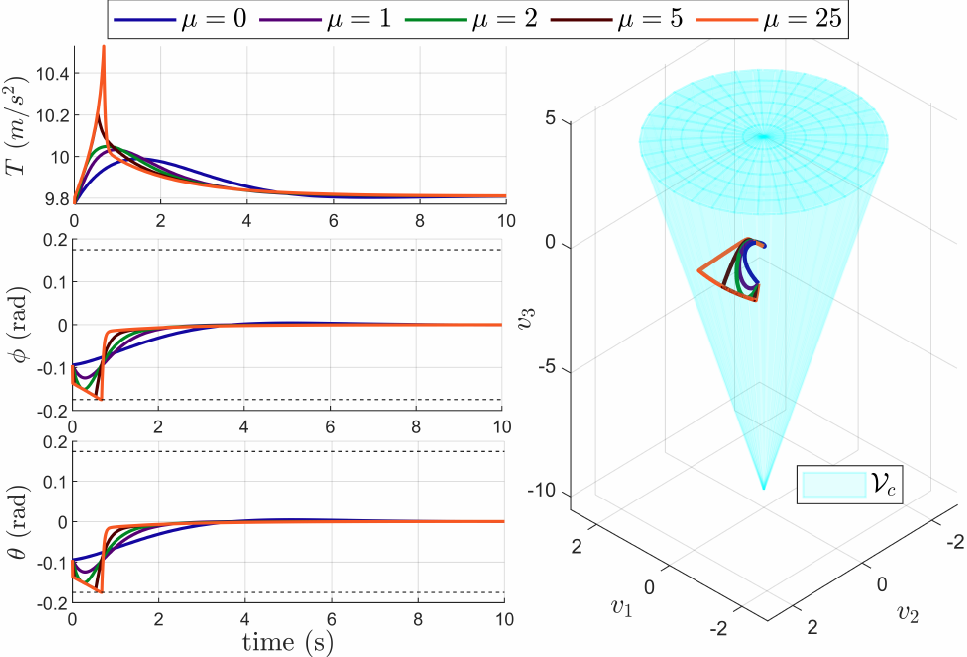}
    \caption{Real input $\bu$ and the input $\bv$ in the new coordinates.}
    \label{fig:adaptive_sim_UV}
    \vspace{-0.5cm}
\end{figure}

In Fig. \ref{fig:adaptive_sim}, with $\mu$ varying from 1 to 5, the convergence time is improved compared to that of the case $\mu=0$ (i.e., without any adaptation). However, in the same figure, the redundancy of excessively high adaptation speed is also highlighted. More specifically, although the performance in the transition state (from 0 to 2 seconds) is ameliorated accordingly to the increase of $\mu$, the convergence times near the steady state (from 2 to 3.5 seconds) become higher. This can be explained  by the ``drifting" saturation effect of the input $\bv$ on the surface of $\mathcal{V}_c$ (see Fig. \ref{fig:adaptive_sim_UV}), leading to unnecessary control effort to bring the system to the equilibrium. 

\begin{table}[htbp]
    \centering
    \caption{Controller's parameters for the quadcopter system}
    \begin{tabular}{|c|c|}
    \hline
    Parameters& Values \\ \hline
       $\umax T$ as in \eqref{eq:constrU}  & $1.45g\approx 14.22 m/s^2$  \\\hline
       $\umax \epsilon$ as in \eqref{eq:constrU}  & $0.1745$ rad (10$^o$)  \\\hline
       $(\gamma_0,V_\infty)$ as in \eqref{eq:adapt_gamma_b},\eqref{eq:thresholdV} & $(1,0.05)$\\\hline
       $\boldsymbol Q$ and $\alpha$ as in \eqref{eq:stabilizingcondi} & $\bbm  2.3148 \bI_3 & -1.3889 \bI_3 \\ -1.3889 \bI_3 & 1.6667 \bI_3\ebm, 1.2$\\\hline
       $\rho$ as in \eqref{eq:BallinVc}, $\varepsilon$ as in \eqref{eq:ellipsoidE}& $2.9019,2.4182$\\\hline
       The bound of $t_\infty$ as in \eqref{eq:bound_tinfty}& $3.2323
$ (s)\\\hline
    \end{tabular}
    \label{tab:para_sim_adpt}
    \vspace{-0.3cm}
\end{table}

Moreover, although the input's constraint satisfaction, its continuity and the system's stability are always guaranteed, the saturation effect with high gain feedback result in the input signal $\bu$ with relatively high rate of change. This  might exceed the physical limit of the system's actuators. Therefore, an overly big adaptation gain $\mu$ is not recommended both for theoretical simulation and practical implementation. Besides, our upper-bound given in \eqref{eq:bound_gamma} for the non-decreasing evolution of $\gamma(t)$ is also validated via the simulation tests, as depicted in Fig. \ref{fig:adaptive_sim}. It can be observed that, due to the increasingly fast transitional behavior with respect to $\mu$, the first inequality of \eqref{eq:bound_gamma} become loosen, hence creating a correspondingly larger gap between $\gamma(t_\infty)$ and the precomputed upper bound as given in \eqref{eq:bound_gamma}.

Next, the control strategy is modified to characterize the robustness of the method in the presence of disturbances.
\subsection{Robustness analysis}
Consider the disturbed model of the quadcopter as follows:
\be 
\dot \bxi = \boldsymbol A\bxi + \boldsymbol B(\boldsymbol R_\psi \boldsymbol f(\bu) - g\boldsymbol e_3) + \boldsymbol E\bw,
\label{eq:nonlinear_dis}
\ee 
where $ \bw$ denotes the disturbance vector which is bounded as $\bw^\top\bw\leq 1$. With \eqref{eq:reformU}, the control problem \eqref{eq:nonlinear_dis} becomes:

\be 
\dot \bxi = \boldsymbol A\bxi +\boldsymbol  B\bv + \boldsymbol E\bw.
\label{eq:linear_dis}
\ee 

Then, our saturated control is adjusted as follows.
\begin{prop}
    Take a box $\mathcal{B}\subset \mathcal{V}_c$ in \eqref{eq:Vc_set} described as:
    \be
\mathcal{B}\triangleq\{\bv=[v_1,v_2,v_3]^\top  :|v_i|\leq \bar v_i, i=1,2,3\}.
\label{eq:box_inVc}
\ee 
Then, 
the ellipsoid
\be 
\mathcal{E}_w=\{\bxi\in\R^6:\bxi^\top \boldsymbol P_w\bxi \leq 1\},
\ee
is robust positively invariant,
under the controller:
\be 
\bv=\text{sat}_\lambda(-\gamma \boldsymbol B^\top \boldsymbol P_w\bxi),\,\forall\gamma\geq 1,
\label{eq:sat_robust_controller}
\ee 
where, for some scalar $\beta >0$, $\boldsymbol P_w\triangleq \boldsymbol Q_w^{-1}$  is found by solving the following LMIs:
\begin{subnumcases}
{\label{eq:robust_cond}}
\bbm 
\bar v_i^2& \row_i(\boldsymbol B^\top) \\
\row_i(\boldsymbol B) & \boldsymbol Q_w
\ebm \succeq 0, i\in\{1,2,3\},& \label{eq:robust_cond_a}\\
\boldsymbol Q_w \boldsymbol A^\top + \boldsymbol A\boldsymbol Q_w-2\boldsymbol B\boldsymbol B^\top +\beta \boldsymbol Q_w+\beta ^{-1}\boldsymbol E\boldsymbol E^\top .
&  \label{eq:robust_cond_b}
\end{subnumcases}
\vspace{0.01cm}
\end{prop}
\begin{proof}
The proof for the invariance of $\mathcal{E}_w$ is threefold. First, we show that with \eqref{eq:robust_cond_a}, the nominal control yields:
\be 
\bv_{nom}\triangleq- \boldsymbol B^\top \boldsymbol P_w\bxi \in \mathcal{B}, \forall\bxi \in \mathcal{E}_w.
\label{eq:vnom}
\ee 
Indeed, inside $\mathcal{E}_w$, each row of $\bv_{nom}$ is bounded as:
\be 
|\row_i(-\boldsymbol B^\top \boldsymbol P_w)\bxi |\leq \sqrt{\row_i(\boldsymbol B^\top \boldsymbol P_w)\boldsymbol Q_w\row_i(\boldsymbol B^\top \boldsymbol P_w)^\top },
\label{eq:box_v_cond}
\ee 
for $i\in\{1,2,3\}$. Then, bounding the right-hand side of \eqref{eq:box_v_cond} with $\bar v_i$, performing the Schur complement and pre-post multiplying the result with $ \begin{bsmallmatrix}
1&\boldsymbol 0_{1\times 6}\\\boldsymbol 0_{6\times 1} & \boldsymbol Q_w
\end{bsmallmatrix}$ respectively yield:
\be 
\begin{aligned}
    &\row_i(\boldsymbol B^\top \boldsymbol P_w)\boldsymbol Q_w\row_i(\boldsymbol B^\top \boldsymbol P_w )^\top\leq \bar v_i^2\\
    &\Leftrightarrow \bbm
\bar v_i^2 & \row_i(\boldsymbol B^\top \boldsymbol P_w\bxi ) \\ 
\row_i(\boldsymbol B^\top \boldsymbol P_w\bxi )^\top & \boldsymbol Q_w^{-1}
\ebm \succeq 0\\ 
&\Leftrightarrow\bbm 
\bar v_i^2& \row_i(\boldsymbol B^\top) \\
\row_i(\boldsymbol B) & \boldsymbol Q_w
\ebm \succeq 0.
\end{aligned}
\ee 
Hence, \eqref{eq:vnom} holds under the condition \eqref{eq:robust_cond_a}. 
Second, by the definition \eqref{eq:def_sat_full}, $\forall
{\bv}\in \mathcal{B}\subset {\mathcal{V}_c} , \gamma \geq 1 ,$ we always have:
\be
\|\text{sat}_\lambda(\gamma {\bv} )\|_2^2
=\|\gamma\lambda^*(\gamma{\bv}){\bv}\|_2^2
\geq 
 \| \eta^*(\bv){\bv}\|_2^2 \geq \| {\bv}\|_2^2,
\ee 
where $\eta^*(\bv) \triangleq\underset{\eta\bv\in\mathcal{B}}{\max \eta}$, which leads to:
\be 
\gamma\lambda^*(\gamma{\bv}) \geq 1, \forall {\bv}\in \mathcal{B},\gamma\geq 1.
\label{eq:condition_gain}
\ee 
Then, finally, by using the property \eqref{eq:condition_gain} for $\bv_{nom}\in\mathcal{B},\forall\bxi\in\mathcal{E}_w$ as in \eqref{eq:vnom}, we can show that:
\begin{align}
     &\boldsymbol Q_w\boldsymbol A_{cl}({\bxi})^\top +\boldsymbol A_{cl}({\bxi})\boldsymbol Q_w + \beta \boldsymbol Q_w +\beta^{-1}\boldsymbol E\boldsymbol E^\top \nonumber\\
     &\preceq \boldsymbol Q_w\boldsymbol A^\top + \boldsymbol A\boldsymbol Q_w - 2\boldsymbol B\boldsymbol B^\top +\beta  \boldsymbol Q_w+\beta ^{-1}\boldsymbol E\boldsymbol E^\top \\ 
     &\quad\quad\quad-(\gamma\lambda^*(-\gamma \boldsymbol B^\top \boldsymbol P_w {\bxi})-1)      
 \boldsymbol P\boldsymbol B\boldsymbol B^\top \boldsymbol P \nonumber\\
 &\preceq \boldsymbol Q_wA^\top + \boldsymbol A\boldsymbol Q_w - 2\boldsymbol B\boldsymbol B^\top +\beta  \boldsymbol Q_w+\beta ^{-1}\boldsymbol E\boldsymbol E^\top\preceq 0\nonumber
\end{align}
with $\boldsymbol A_{cl}({\bxi})\triangleq\left(\boldsymbol A-\boldsymbol B\gamma\lambda^*(-\gamma \boldsymbol B^\top \boldsymbol P_w {\bxi})\boldsymbol B^\top \boldsymbol P \right)$.
Namely, with \eqref{eq:robust_cond}, and the controller \eqref{eq:sat_robust_controller}, the closed-loop dynamics:
\be 
\dot \bxi =\boldsymbol A\bxi +\boldsymbol B\text{sat}_\lambda(-\gamma \boldsymbol B^\top \boldsymbol P_w\bxi) + \boldsymbol E\bw 
\ee 
satisfy the invariance condition given in Proposition \ref{prop:invariance_robust}.
\end{proof}

\textit{Simulation study:} Let us analyze the procedure with two different choices of $\gamma$ for the stabilization starting from the boundary points of $\mathcal{E}_w$. The disturbance will be simulated by the bounded wind on $x,y$ axis $\bw\triangleq[w_x,w_y,0,0,0,0]^\top$. Simulation specifications are given in TABLE \ref{tab:para_sim_robust}.
\begin{figure}[htbp]
    \centering
    \includegraphics[scale=0.475]{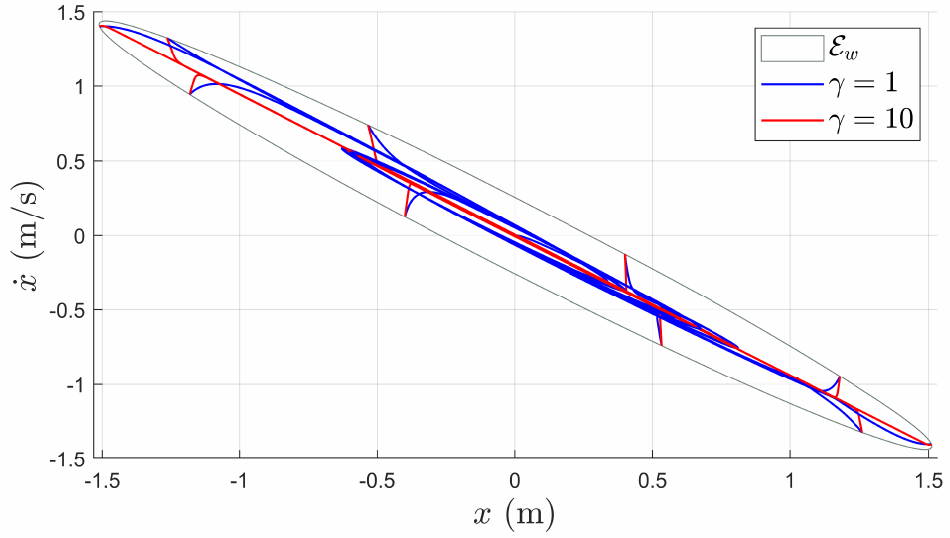}
    \caption{The system trajectories from various initial points, subject to disturbances (projected in ($x, \dot x$) subspace)}
    \label{fig:robust_sim_xxdot}
\end{figure}

Via the simulation, the invariance is validated for both gains $\gamma=1$ and $\gamma=10$. Besides, one noticeable difference between the two choices of gains is that the system's trajectory appears to be more compressed with the latter gain towards the surface of $\boldsymbol B^\top \boldsymbol P \bxi = 0$ (See Fig. \ref{fig:robust_sim_xxdot}). This behavior can be explained by the more aggressive control action generated locally near the surface.
\begin{figure}[htbp]
    \centering
\includegraphics[width=0.45\textwidth]{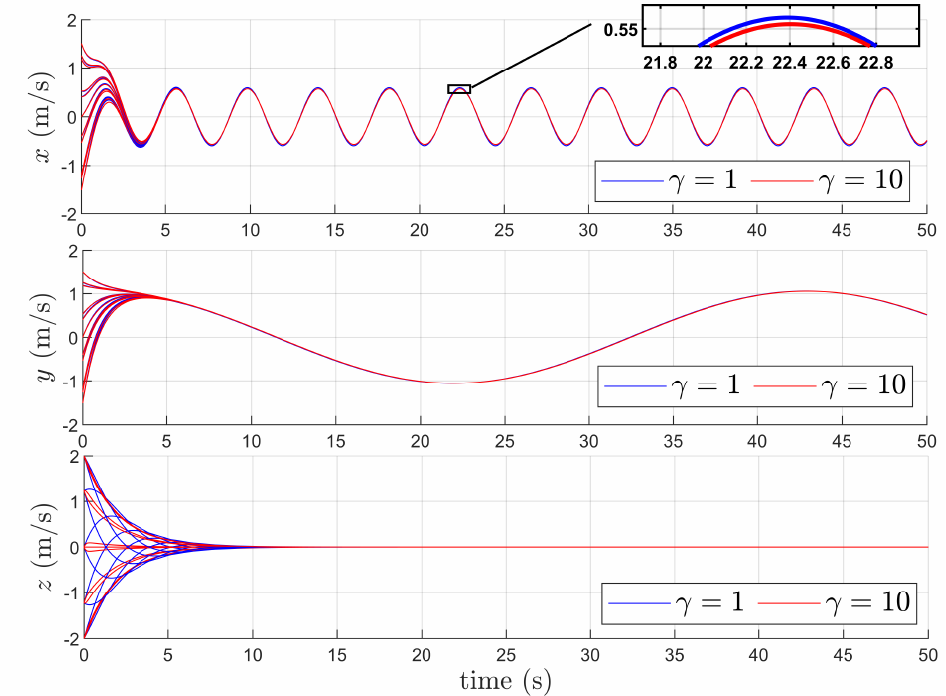}
    \caption{Real input $\bu$ and the new input $\bv$ simulated with disturbances}
    \label{fig:robust_sim_xyz}
    \vspace{-0.35cm}
\end{figure}

 Furthermore, since no disturbance counter-measures (e.g., disturbance estimators) were taken into account by the controller, although the errors are proven to be bounded, few improvements can be observed for both the nominal gain $\gamma=1$ and the high gain $\gamma=10$ (See Fig. \ref{fig:robust_sim_xyz}).

\begin{rem}
    The design procedure can be easily adjusted for the \textit{trajectory tracking} problem as follows. First, we assume that the reference trajectory for the linearized system
    satisfies the dynamics constraints:
    
    \be 
        \dot\bxi^\mathrm{ref}=\boldsymbol A\bxi^\mathrm{ref}+\boldsymbol B\bv^\mathrm{ref};\;
        \bv^\mathrm{ref}\in \mathcal{V}^\mathrm{ref}\subset \mathcal{V}_c 
        \label{eq:ref_sys}
    \ee 
 where $\bxi^\mathrm{ref},\bv^\mathrm{ref}$ denote, respectively, the reference for the state $\bxi$ and input $\bv$; $\mathcal{V}^\mathrm{ref}$ is a polytope subset of $\mathcal{V}_c$ entirely wrapping $\bv^\mathrm{ref}$. Then, the error dynamics yields:
 \be 
\dot{\Tilde{\bxi}}=\boldsymbol A\Tilde{\bxi}+\boldsymbol B\Tilde{\bv} + \boldsymbol E\bw, \text{ subject to: } \Tilde{\bv}\in \Tilde{\mathcal{V}}
\label{eq:constr_track_error}
 \ee 
 with $\Tilde{\bxi}\triangleq\bxi-\bxi^\mathrm{ref}, \Tilde{\bv}\triangleq\bv-\bv^\mathrm{ref}$ and $\Tilde{\mathcal{V}}=\mathcal{V}_c\ominus \mathcal{V}^\mathrm{ref}$.
In this fashion, the trajectory tracking can be achieved by applying the proposed stabilizing controller for system \eqref{eq:constr_track_error}. To simplify the computation of $\Tilde{\mathcal{V}}$, $\mathcal{V}_c$ can be arbitrarily tightly approximated as in \cite{thinhECC23} by a polytope. Moreover, assume $\Tilde{\mathcal{V}}$ is given in half-space form:
 \be 
\Tilde{\mathcal{V}}=\{\bv: \boldsymbol a_{\tilde v,i}^\top \bv\leq b_{\tilde v,i}, b_{\tilde v,i}> 0, i\in \{1,...,N_{\Tilde{\mathcal{V}}}\}\},
 \ee 
 Then, the saturation function over the set $\Tilde{\mathcal{V}}$ is described as:
 \be
        {\text{sat}}_{ \Tilde\lambda} (\tilde\bv) \text{=}\Tilde\lambda^*(\tilde\bv)\tilde\bv ,        
\text{with } \Tilde\lambda^*(\tilde\bv)\text{=}
 \begin{cases}
            1 &\text{ if } \tilde\bv\in \Tilde{\mathcal{V}},\\
 \underset{{
\Tilde\lambda \bv \in \Tilde{\mathcal{V}}}}{ \max }\Tilde\lambda
 &\text{ if } \tilde\bv\notin \Tilde{\mathcal{V}},
         \end{cases}
\label{eq:LP_prob_sat}
 \ee 
 and can be explicitly computed, with $i\in \{1,...,N_{\Tilde{\mathcal{V}}}\}$, as:
 \be 
         \Tilde{\text{sat}}_\lambda (\tilde\bv) \text{=}\begin{cases}
            \tilde\bv &\text{ if } \tilde\bv\in \Tilde{\mathcal{V}},\\
\Tilde{\bv} \big(\underset{
\boldsymbol a_{\tilde v,i}^\top \tilde\bv>0
}
{ \min }
b_{\tilde v,i}/(\boldsymbol a_{\tilde v,i}^\top \tilde\bv)\big)
 &\text{ if } \tilde\bv\notin \Tilde{\mathcal{V}}.
         \end{cases}
         \label{eq:solution_LP}
 \ee 
Namely, the formula \eqref{eq:solution_LP} is the solution of the linear programming problem employed in \eqref{eq:LP_prob_sat}.
\end{rem}

\begin{table}[tb]
    \centering
    \caption{Simulation parameters for robustness analysis}
    \begin{tabular}{|c|c|}
    \hline
    Parameters& Values \\ \hline
       $\umax T$, $\umax \epsilon$  & $2g,0.698$ (rad)  \\\hline
       $Q_w$ and $\beta$ in \eqref{eq:robust_cond} & $\bbm   \text{diag}(2.29,2.29, 4.42) & -2.14 \bI_3 \\ -2.14\bI_3 & 2.07 \bI_3\ebm,     0.9668$\\\hline
       $w_x,w_y$  & $\sin({1.5t+\pi/8}),\cos({0.15})$ (m/s)  \\\hline
       $\bar v_1,\bar v_2,\bar v_3$  in \eqref{eq:box_inVc}& $    3.8804 ,   3.8804,    3.2700
$ \\\hline
    \end{tabular}
    \label{tab:para_sim_robust}
    \vspace{-0.5cm}
\end{table}

\section{Experimental validation}
\label{sec:expval}
For the experiment, the continuous dynamics of $\gamma(t)$ as in \eqref{eq:adapt_gamma_b} will be discretized via the forward Euler method together with the sampling time of $0.1$ (s). 
The real input $\bu$ of the system is calculated on a computer and sent to the quadcopter via a long range 2.4 GHz USB dongle together with the desired yaw angle $\psi$ assigned as $0$.
Next, to highlight the controller's applicability, we first present the experiment tests applied, then discussions will follow. 
\subsection{Experiment setups}
In this section, we carry out the following scenarios.

\begin{itemize}
    \item \textit{Test 1}: the real system behavior under different choices of adaptive gain varying from $\mu=1$ to $\mu=5$ is studied to provide insights on the proper choice of such parameter. The control objective is to track a static point located at $\bxi=[0.6,0.6,0.8,0,0,0]^\top$ starting from the initial point $\bxi(0)=[-0.6,-0.6,0,0,0,0]^\top$.
     \item \textit{Test 2}: To underline the computational advantage, three nanodrones are utilized to track their assigned trajectories in a centralized manner with the proposed scheme. The scenario describes the swapping position of the drones with three anchor waypoints: $\{[0,-6,1],[6,6,1],[-6,6,1]\}\times 10$ (cm) while the references are generated with B-spline trajectory generation \cite{prodan2019necessary}. The three drones and their corresponding reference are labeled as Drone $r$ and Ref. $r$ with $r\in\{1,2,3\}$.
\end{itemize}
The matrix $ \boldsymbol{P}$ and $\varepsilon$ forming the invariant set $\mathcal{E}$ are fixed as: $\boldsymbol P=\bbm 0.98\bI_3 & 0.78\bI_3 \\0.78\bI_3 &1.25\bI_3\ebm$ and $2.3215$, respectively, with the procedure in Proposition \ref{prop:saturatedcontrol} and the input limit in TABLE \ref{tab:para_sim_adpt}.

\subsection{Experimental results and discussion}
\begin{table}[tb]
\vspace{-0.15cm}
\centering
\caption{Numerical parameters \& results for test 2}
\begin{tabular}{|l|lll|}
\hline
 & \multicolumn{1}{l|}{Drone 1} & \multicolumn{1}{l|}{Drone 2} &Drone 3  \\ \hline
RMS tracking errors (m) & \multicolumn{1}{l|}{$0.0703$} & \multicolumn{1}{l|}{$0.0826$} & $ 0.0602$ \\ \hline
Average computation time& \multicolumn{3}{l|}{$1.4809$ (ms)}                            \\ \hline
 $(\gamma_0,\mu, V_\infty)$ & \multicolumn{3}{l|}{$(1.0,2.0,0.05)$}                            \\ \hline
\end{tabular}
\label{tab:expres}
\end{table}

\begin{figure}[htb]
    \centering
    \includegraphics[width=0.475\textwidth]{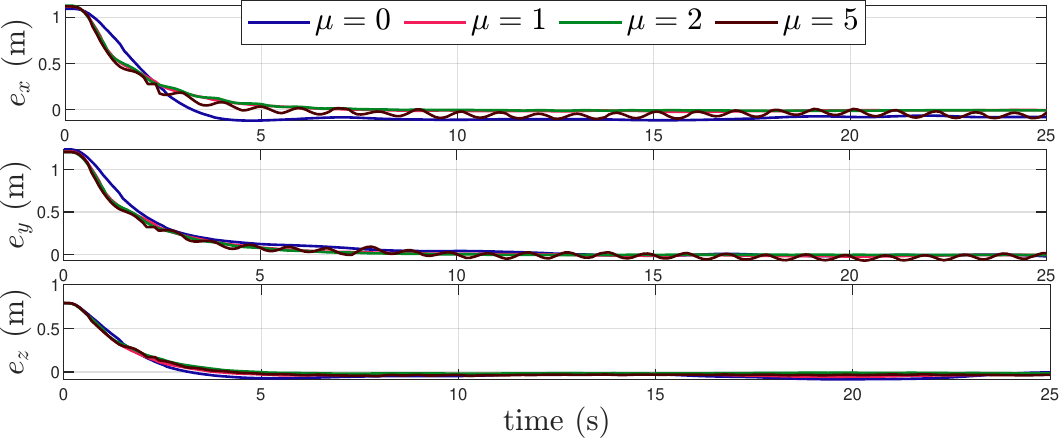}
    \caption{Set-point tracking errors ($e_q=q-q^{\mathrm{ref}},q\in\{x,y,z\}$) via different choices of adaptive gain $\mu$ (Test 1).}
    \label{fig:setpts_4mu}
    \vspace{-0.25cm}
\end{figure}

\begin{figure}[htb]
    \centering
    \includegraphics[width=0.475\textwidth]{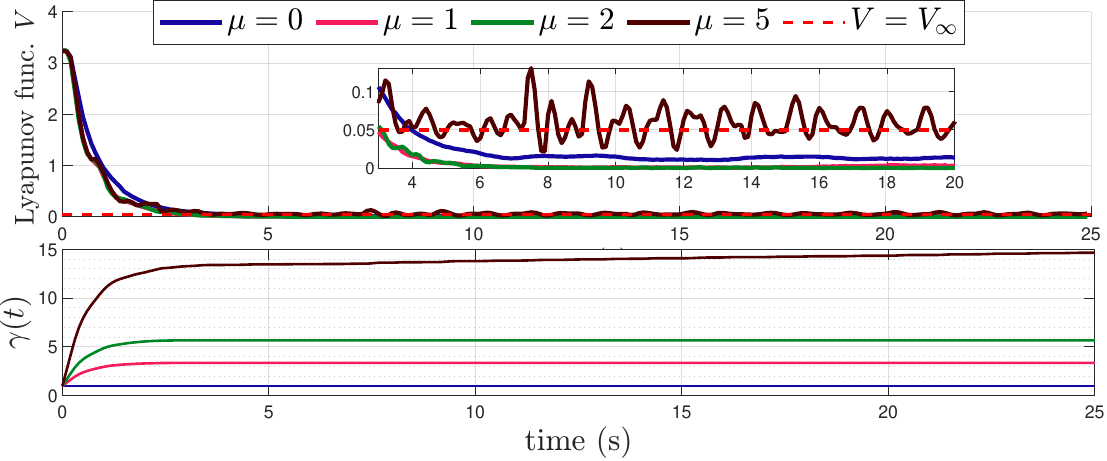}
    \caption{Values of the Lyapunov function (top) and the evolution of $\gamma(t)$ (bottom) with different adaptive rate $\mu$.}
    \label{fig:Lya_4mu}
    \vspace{-0.3cm}
\end{figure}

\begin{figure}[htb]
    \centering
    \includegraphics[scale=0.5]{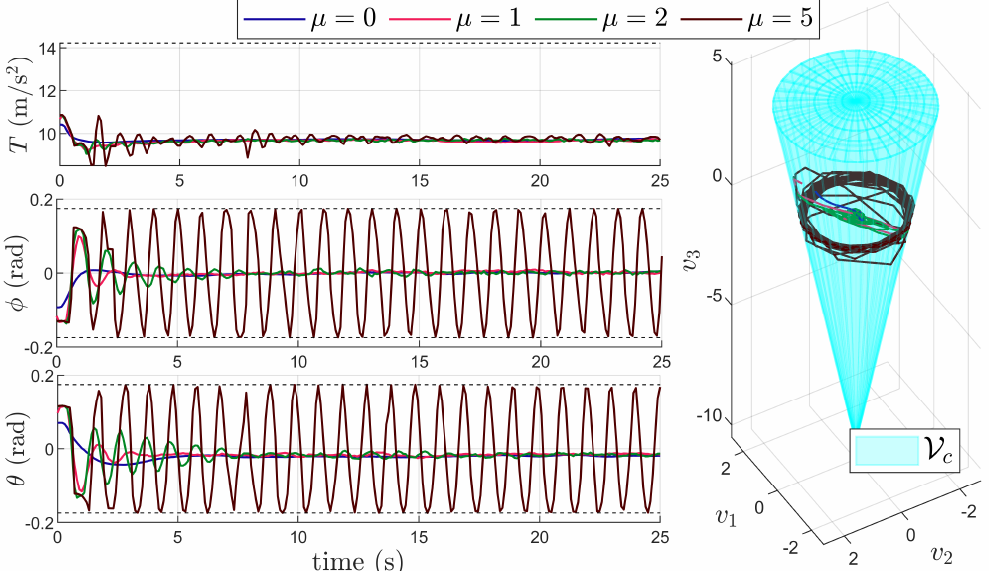}
    \caption{Input $\bu$ and $\bv$ via different choices of adaptive gain $\mu$.}
    \label{fig:inputs_exp}
    \vspace{-0.35cm}
\end{figure}

In general, the stabilization effect and constraint satisfaction guarantee are validated when the quadcopter arrives to the reference point with the input $\bu$ remaining correctly bounded in $\mathcal{U}$ (See Fig. \ref{fig:inputs_exp}). However, as in Fig. \ref{fig:Lya_4mu}, although the stability is secured for all cases, the Lyapunov function of $\mu=5$ is not confined under the threshold $V=V_\infty$ as in \eqref{eq:thresholdV}. This oscillation can also be observed in the tracking error (Fig. \ref{fig:setpts_4mu}). The behavior was caused by the high 
adaptation speed $\mu$ abruptly increasing the control gain $\gamma(t)$ to a large value, which excites the neglected high frequency dynamics (e.g., rotation dynamics). Furthermore, this high gain of $\mu$ also
prevents $\gamma(t)$ from converging (Fig. \ref{fig:Lya_4mu}). Hence, although in the examined scenarios, the stability is maintained, an unreasonably high adaptation speed $\mu$ is not recommended. 

On the contrary, with a moderate choice of $\mu$ ($\mu=1$ or $2$), the steady state tracking error is removed (See Fig. \ref{fig:setpts_4mu}). Furthermore, the appropriate value of $\gamma(t)$ also converges to a fixed value (Fig. \ref{fig:Lya_4mu}), solving the problem of choosing the gain $\gamma $ for the saturated control described in \eqref{eq:controller_sat}.

\begin{figure}[htbp]
    \centering
    \includegraphics[scale=0.5]{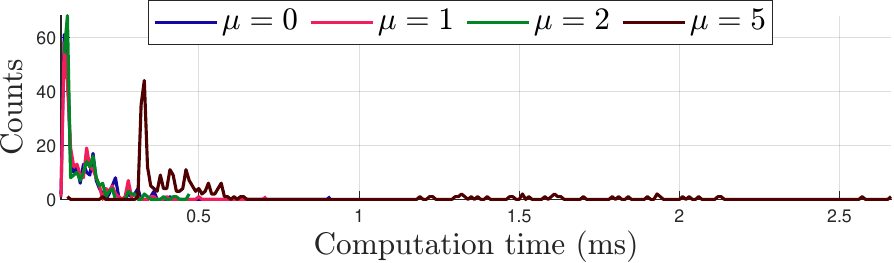}
    \caption{Computation time histogram with Test 1.}
    \label{fig:comptime_histo_4mu}
    \vspace{-0.15cm}
\end{figure}

Besides, thanks to the explicit solution of the saturated function as given in \eqref{eq:KKT_M}, the computation time remains under 3 ms (See Fig. \ref{fig:comptime_histo_4mu}) which is relatively small compared to the existing results of nonlinear, optimization-based controller providing the same guarantees \cite{nguyen2020stability,mueller2013model,nguyen2019stabilizing}. This advantage is further highlighted with the reliable control of three quadcopters in a centralized manner in Test 2 with less than $2$ ms in the average of computation time (TABLE \ref{tab:expres}) with under 10 cm of root-mean-square (RMS) tracking errors. 

\begin{figure}[htbp]
    \centering
    \includegraphics[scale=0.5]{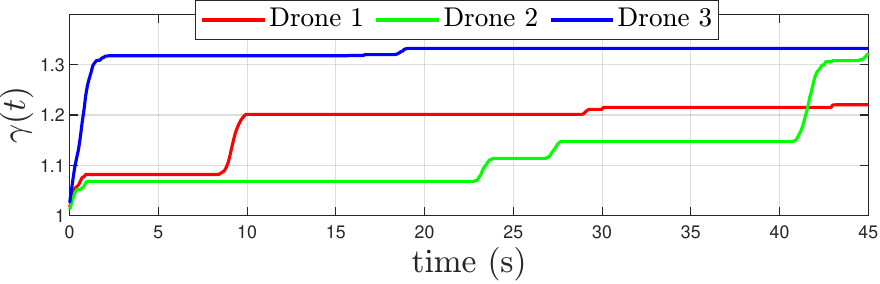}
    \caption{The evolution of $\gamma(t)$ for the three quadcopters}
    \label{fig:gamma_3drones}
\end{figure}

However, from Fig. \ref{fig:gamma_3drones}, although the $\gamma(t)$ value of Drone 3 has the tendency to converge, that of Drone 1 and Drone 2 increases correspondingly to the vehicle's deviation from the reference due to disturbances. This phenomenon, as a future concern, complicates
both the problem of disturbance rejection for the strategy and the adjustment for
the non-monotonicity of the gain $\gamma(t)$ in the presence of disturbances so that input over-exploitation can be avoided.
\begin{figure}[htbp]
    \centering
    \includegraphics[width=0.465\textwidth]{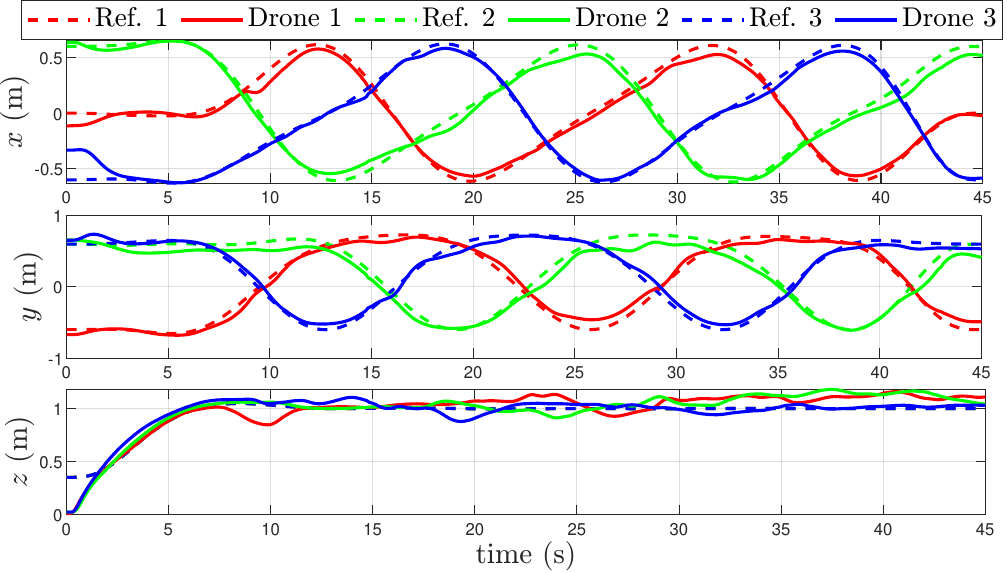}
    \caption{Position swapping with three Crazyflies (Test 2)}
    \label{fig:trajtrack_3drones}
    \vspace{-0.2cm}
\end{figure}

Finally, thanks to the system's linear representation in the flat output space, a computationally simple, yet, effective controller with stability and constraint satisfaction can be developed as opposed to other implicit optimization-based schemes \cite{thinhECC23,nguyen2020stability}. This not only leads to the simplification of control design, but also opens the door to possible control synthesis for cooperative scenarios.


\section{Conclusion}
\label{sec:conclude}
This paper presented an adaptive saturated control scheme with the flatness-based variable change. More specifically, in the flat output space, the system was represented by a linear dynamics at the cost of convoluted input constraints. By exploiting their convexity, a non-standard saturated scheme with adaptive gain was presented. The main features of the proposed scheme were the stability guarantees, constraints satisfaction, gain adaptivity,
and robustness characterization. These were further 
validated via both simulations and experiments. Future work will concentrate on the design of a robust saturated controller applied for multiple drones control.

\end{document}